# Alterations in Structural Correlation Networks with Prior Concussion in Collision-Sport Athletes

Muhammad Usman Sadiq[1], Diana Svaldi[2], Trey Shenk[1], Evan Breedlove[2], Victoria Poole[2], Greg Tamer[2], Kausar Abbas[1], Thomas Talavage[1]

[1]School of Electrical & Computer Engineering, Purdue University, West Lafayette, IN,
[2]Weldon School of Biomedical Engineering, Purdue University, West Lafayette, IN

## Abstract

Several studies have used structural correlation networks, derived from anatomical covariance of brain regions, to analyze neurologic changes associated with multiple sclerosis, schizophrenia and breast cancer [1][2]. Graph-theoretical analyses of human brain structural networks have consistently shown the characteristic of small-worldness that reflects a network with both high segregation and high integration. A large neuroimaging literature on football players, with and without history of concussion, has shown both functional and anatomical changes. Here we use graph-based topological properties of anatomical correlation networks to study the effect of prior concussion in collision-sport athletes. 40 high school collision-sport athletes (23 male football, 17 female soccer; CSA) *without* self-reported history of concussion (HOC-), 18 athletes (13 male football, 5 female soccer) *with* self-reported history of concussion (HOC+) and 24 healthy controls (19 male, 5 female; CN) participated in imaging sessions before the beginning of a competition season. The extracted residual volumes for each group were used for building the correlation networks and their small-worldness, $\sigma$, is calculated. The small-worldness of CSA *without* prior history of concussion, $\sigma_{HOC-}$, is significantly greater than that of controls, $\sigma_{CN}$. CSA *with* prior history have significantly higher (vs. 95% confidence interval) small-worldness compared to HOC+, over a range of network densities. The longer path lengths in HOC+ group could indicate disrupted neuronal integration relative to healthy controls.

## 1. Introduction and literature review

American football, the most played sport in the United States, has drawn interest and concern related to short-term and long-term effects of concussions sustained on the field [3] [4][5]. Although the true number is speculated to be more, the estimated number of concussion incidences each year stands at about 1.6-3.8 millions[6]. Moreover, research has shown that mild head injury can accumulate to a more severe injury, and athletes with undiagnosed concussions are at a higher risk of long-term neurological damage [7][8].

Anatomical images are used to study anatomical effects - volume and surface thickness changes - accompanying various disorders[9][10]. Given the high incidence of sports-related traumatic brain injury and the accompanying functional changes, the question of whether exposure to head impact in football results in short-term or long-term neuroanatomical changes becomes relevant. Brain atrophy has been studied in the context of traumatic brain injury before, and evidence points to regional white matter and gray matter changes following mild traumatic brain injury [11][12][13], with recent examples of functional and structural impairments in collision-sports athletes[14][15][16]. Hofman et al showed a loss of total brain volume after mild TBI, while

Arciniegas et al demonstrated reduced hippocampal volume following TBI[17] . Cortical atrophy and changes in lesion volume and ventricular size have also been associated with TBI[18][19].
One key limitation of studying atrophy in individual regions of interest (ROIs) following TBI is that with diffuse nature of the injury, the damage sustained is a strong function of location. This implies that even when the head impact may not be causing individual damage to a specific ROI, it may change the overall topology, functional and structural connectivity and information flow. Due to this, a graph-based anatomical connectivity analysis is often performed along with study of voxel-based or ROI-based alterations.

The idea of anatomical connectivity between two regions based on statistically significant association in cortical thickness or density has been explored in a few studies [20][21]. Two regions are considered anatomically correlated if their size co-varies across subjects. It has been hypothesized that the covariation of morphological features in related cortical regions can arise from a mixture of mutual influences, experience-related changes and genetics [22][23][24]. More support for such coupling comes from a study [25] where different elements of the human visual system showed covariance across individuals, as well as evidence of enlarged sensorimotor, premotor and parietal areas in trained musicians[26]. Since the quantitative definition of graph theoretical properties for structural correlation networks, several studies have attempted to study their properties [27][28]. In particular, changing patterns in structural correlation networks have been associated with normal aging, multiple sclerosis, schizophrenia, breast cancer and epilepsy [29][2][30][31]

Despite examples of where correlation networks have been used to study atrophies, the relationship between morphological correlation and functional or structural connectivity is not very clear. Although functional changes can indicate structural and morphological changes, and vice versa, the evidence connecting the two does not always suggest a very direction correlation. For example, although initial study showed that 15 pairs of highly correlated regions were associated with previously identified WM fiber tracts[20], a subsequent study suggested that only 35-40% of morphological correlations are associated with underlying WM connections [32]. Therefore, due to evidence of cognitive and regional GM density changes in collegiate football athletes with history of concussion [16], the question of whether their structural correlation networks differ or not becomes important.

Recent graph-theoretical analyses of the human brain structural networks have consistently shown the characteristic of small-worldness, a property that defines a network with both high segregation and integration. Traditional univariate neuroimaging analyses of football players, with and without history of concussion, have shown both functional and anatomical changes. We, therefore, attempt to analyze graph-based topological properties of anatomical correlation networks. One advantage of graph-based approach over univariate analysis is that it can directly test the topological differences between graph properties, providing more insight into cognitive status and connectivity.

## 2. Methods

### 2.1 Subjects

40 high school collision-sport athletes (23 male football, 17 female soccer athletes) *without* history of concussion (HOC- ages:14-18, average = 16.4), 18 athletes (13 male football,5 female soccer ) *with* self-reported history of concussion (HOC+ ages:14-18, average = 16.7) were scanned from 3 high schools across three competitions seasons (Season 3: 2011-12, Season 4: 2012-13 and Season 6: 2014-15). Additionally, 25 healthy controls (ages: 14-18, average = 15.7) participating in non-collision sports (19 male, 5 female) from 2 different local schools, were scanned across 2 competition seasons. All subjects participated in imaging sessions 1-2 months after the end of a competition season. Note that all players with history of concussion (HOC+) also self-reported at least one of the symptoms from the Concussion Symptom Inventory at the time of imaging.

## 2.2 Processing

Anatomical scans were reconstructed using the FreeSurfer image analysis suite [33]. FreeSurfer's processing pipeline has been shown in Figure 3.1 [33][34], where the typical steps include motion correction, skull-stripping, volume transformation and labelling followed by surface segmentation and registration. Volumetric segmentation and parcellation were followed by additional scripts to segment hippocampal sub-fields [35] .

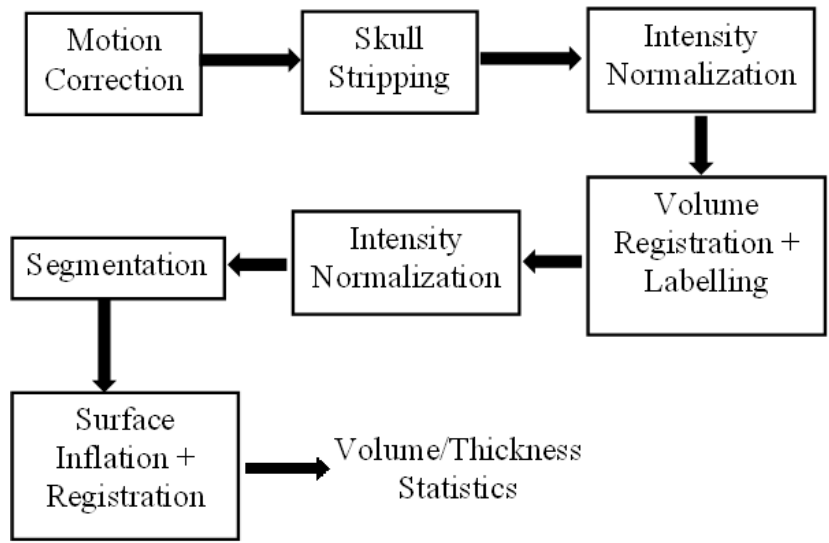

Figure 1: FreeSurfer anatomical processing pipeline

Each reconstructed anatomical session was subject to the FreeSurfer quality check protocol[36]. The protocol followed the standard steps of pial surface edits, white matter edits, topology edits, checking the skull-strip, intensity normalization and Talairach transform. The subjects were re-processed after edits and subsequently considered permanent failure in case the reprocessing did not make the requisite corrections.

## 2.3 Analysis

The extracted residual volumes for each group were used for building the correlation networks. For each group, a 68 x 68 Adjacency matrix A was acquired, where $A_{ij}$ is the Pearson correlation coefficient between regions $i$ and $j$. For the purpose of translating this matrix into an undirected graph, these coefficients need to be compared with a threshold, which tends to controls the network characteristics and the subsequent analysis[1]. We, therefore, use the method described in [1] which thresholds the adjacency matrix at a range of network densities, and compares the topological characteristics.

### 2.3.1 Graph metrics for comparison

Human brain networks can be identified distinctly from random networks of the same order by key network characteristics such as small-worldness. The property of small-worldness of a network depends upon its two metrics: the characteristic path length $L$ and clustering coefficient $C$. Clustering coefficient is a measure of number of edges in any neighborhood in a network, and quantifies the amount of segregation a network possesses. The characteristic path length $L$ is the average shortest path length between two nodes in a network, and is a measure of its integration. For small-worldness, we know that $\gamma = C/C^{rand} > 1$ and $\lambda = L/L^{rand} \sim 1$. These conditions can be combined as small-worldness $\sigma = \frac{\gamma}{\lambda} > 1$ [37]. Small world networks are therefore characterized by high segmentation and high integration.

Based on these definitions, He et al[20] found that the mean clustering coefficient of the gray matter network in healthy humans was about twice that of a random network, whereas path length is comparable. This results in $\gamma = 2.36$, $\lambda = 1.15$, and $\sigma = 2.04$ for human cortical anatomical network, and $\gamma = 2.08$, $\lambda = 1.09$, $\sigma = 1.91$ for human brain functional network, both reflecting high degree of small-worldness.

We perform pairwise comparisons between HOC+, HOC- and Controls graphs. To calculate small-worldness $\sigma$ for our networks, the clustering coefficient and path length need to be compared with the corresponding values of a random graph with the same number of nodes, edges and degree distribution. We use Graph Analysis Toolbox to generate 1000 such networks to determine small-worldness $\sigma$ for each group [38]. In order to test the significance of the difference between topological properties of each pair, we need to generate random graphs with similar topological properties. We use the topology randomization algorithm proposed [39] that preserves the number of nodes, degree and distribution, and randomly reassigns each subjects volume data to the opposite group to generate 1000 randomized graphs. The difference between characteristics of randomized graphs thus generated provides the difference under the null hypothesis. The difference between the two groups being compared is appropriately positioned in the population of graphs, according to their percentile ranks.

# 3. Results

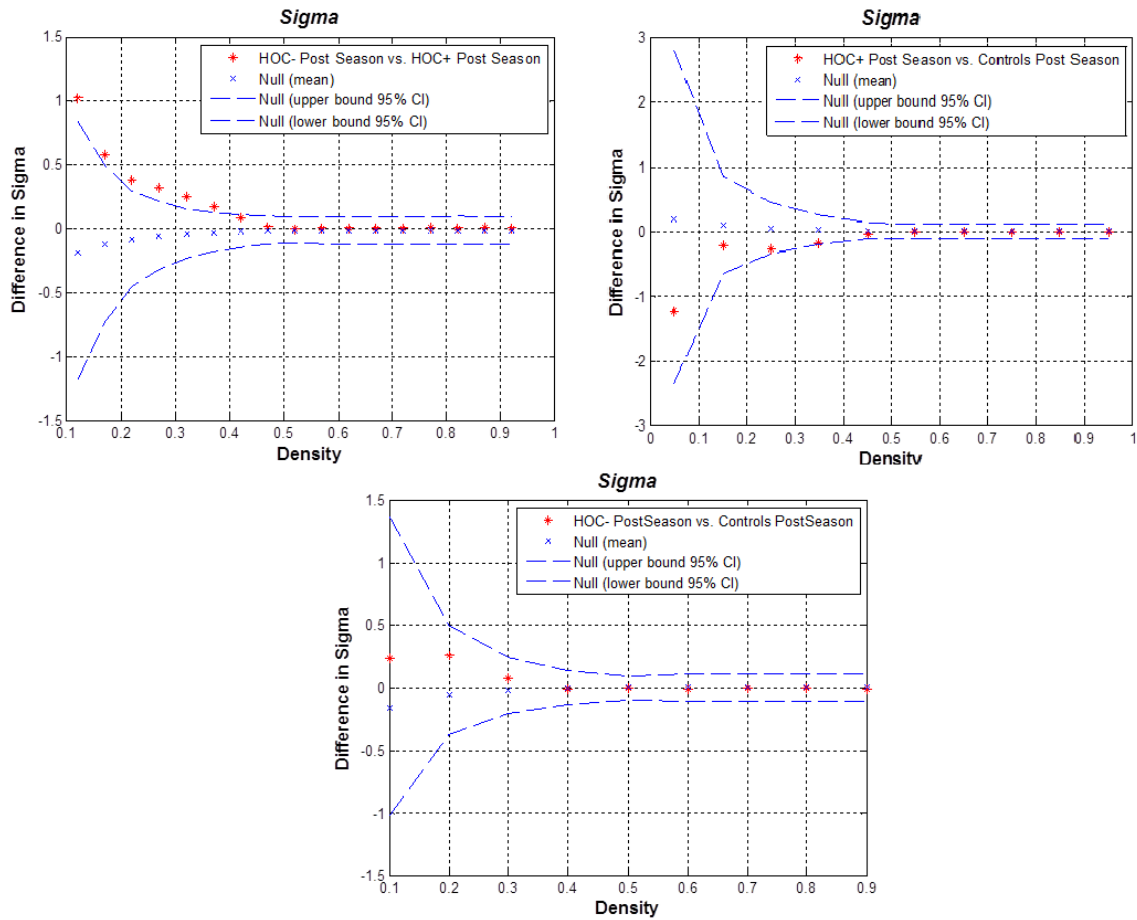

Figure 2: Comparison of small-worldness of Controls, HOC+ and HOC- groups (a-c)

The plot in Fig 2 compares the small-worldness σ for HOC+ and HOC- networks for a range of densities. To test the statistical significance of the difference between σ, we use the topology randomization algorithm in GAT to generate 1000 randomized graphs. The difference between σ of randomized graphs thus generated provides the difference under the null hypothesis.

We can see that the $\sigma_{HOC-}$ is significantly higher than $\sigma_{HOC+}$ using a 95% confidence interval, for a range of densities (0.1-0.4), as shown in Fig 2(a). The plots in Fig 2 (b) and (c) show the σ difference of HOC+ and HOC- graphs with those of controls. We can see that $\sigma_{HOC-} > \sigma_{CN}$, and $\sigma_{HOC+} < \sigma_{CN}$, for the mentioned range of densities (0.1-0.4), although the difference lies within the 95% confidence intervals.

The higher σ in HOC- group can be traced to lower path lengths and lower clustering than the HOC+ group, over the range of densities, as shown in Fig 3.

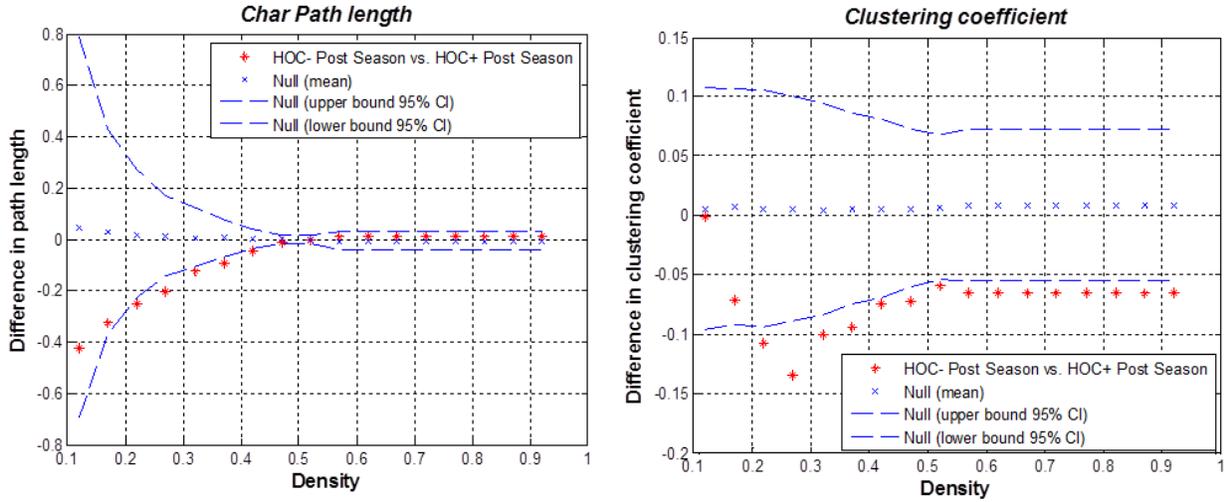

Figure 3: Comparison of network properties of HOC- and HOC+ groups (a) Characteristic Path Length and (b) Clustering Coefficient

## 4. Discussion

In human brain networks, short path lengths are attributed to rapid transfers of information between remote regions that form the basis of cognition [30],[40]. The longer path lengths in HOC+ group can therefore indicate disrupted neuronal integration than healthy controls. The shorter path lengths in HOC- group, on the other hand, reflects better integration and rapid information flow. The HOC+ network also exhibited higher clustering, indicating strong localization. The two factors of clustering and path length, when combined in small-worldness, result in a significant difference between $\sigma_{HOC+}$ and $\sigma_{HOC-}$. It should be noted that even though $\sigma_{HOC+}$ is not significantly different from $\sigma_{controls}$, the notable difference could mean potential disturbance in integration pattern of the network in HOC+ group.

### 4.1 Regional network comparison

It has been shown in literature [41] that regional network characteristics might exhibit change without affecting the overall properties of the network significantly. The regional comparison also helps localize the differences between networks and provide more insight into the implications. We show the association matrices for HOC+ and HOC- groups in Fig 4, and the corresponding regional differences between clustering coefficients in Fig 5.

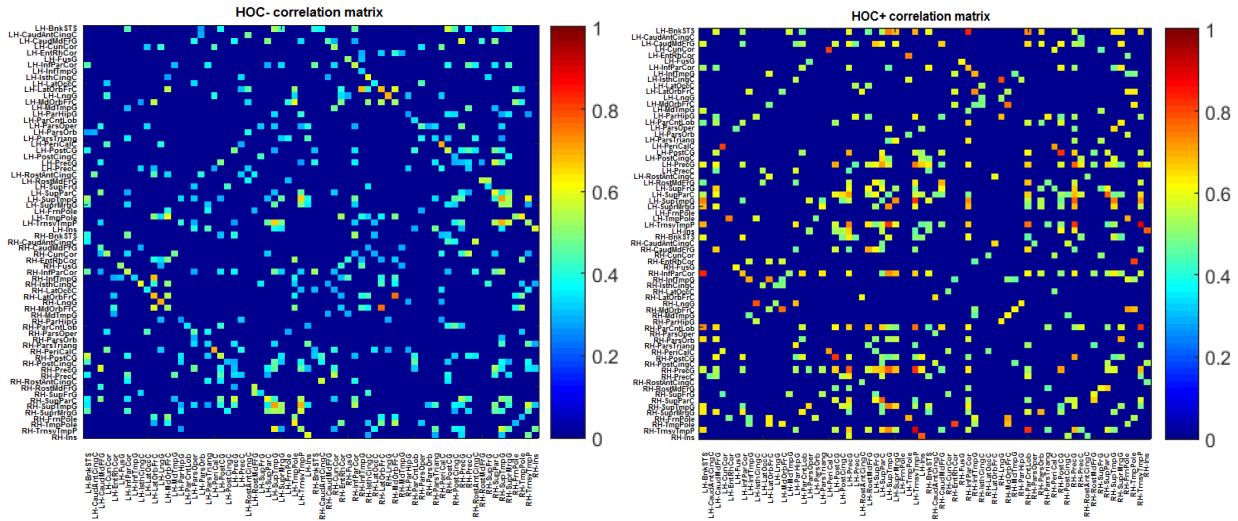

Figure 4: Correlation matrices of HOC- and HOC+ groups thresholded at minimum density of full-connectivity

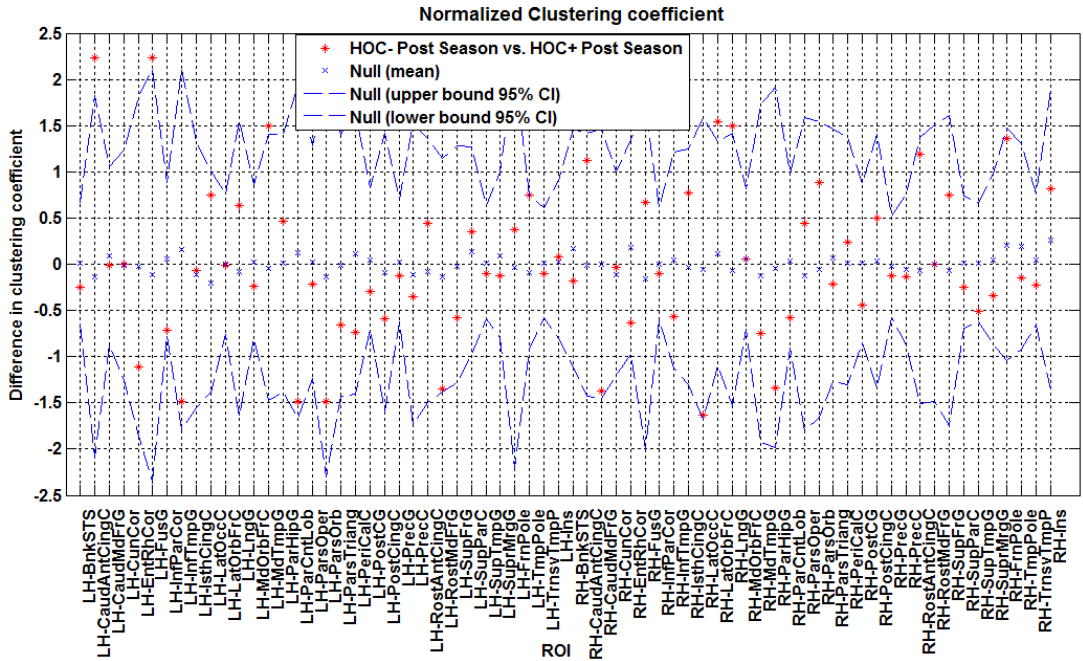

Figure 5: Regional differences between clustering coefficients of HOC- and HOC+ groups

While we can readily see from Fig. 5 that $C_{HOC-} < C_{HOC+}$ for most ROIs, we can also observe that most of this difference comes from LH Fusiform Gyrus, LH Inferior Parietal Cortex, LH Paracentral Lobule, LH Rostral Middle Frontal Gyrus, RH Caudal Middle Frontal Gyrus and RH-Lateral Occipital Cortex. From Fig 5, we can also readily see a more diffuse correlation pattern in HOC- Association matrix than in HOC+, corroborating the earlier finding that HOC- networks exhibit less clustering and higher integration.

## 5. Conclusions

The study shows differences in structural correlation networks between collision-sport athletes with and without history of concussion and non collision-sport controls. Our findings show that strucural networks of athletes with history of concussion (HOC+) possess significantly lower small-worldness than those without prior concussion history (HOC-). The loss of small-wordlness can be traced to significantly longer path lengths and higher clustering in concussed athletes. Regional analysis indicates most significant differences in clustering in Fusiform Gyrus, Inferior Parietal and Rostral Middle Frontal Cortex. Longer path lengths are attributed to slow transfers of information between remote regions that form the basis of cognition [30][40]. These preliminary findings suggest (with caveats related to support from other direct connectivity measures [30] that differences in anatomical correlation based graphs may permit detection and characterization of prior concussive head injury, with implications for subsequent sub-concussive injury.